\documentclass[prb,aps,preprint,superscriptaddress]{revtex4-1}
\usepackage[dvips]{graphicx}
\usepackage{rotating}
\begin{document}
\title{Development of half metallicity within mixed magnetic phase of Cu$_{1-x}$Co$_x$MnSb alloy}

\author{Abhisek Bandyopadhyay}
\email{These two authors have contributed equally}
\affiliation{Department of Materials Science, Indian Association for the Cultivation of Science, 2A \& 2B Raja S. C. Mullick Road, Jadavpur, Kolkata 700 032, India}
\author{Swarup Kumar Neogi}
\email{These two authors have contributed equally}
\affiliation{Department of Materials Science, Indian Association for the Cultivation of Science, 2A \& 2B Raja S. C. Mullick Road, Jadavpur, Kolkata 700 032, India}
\author{Atanu Paul}
\affiliation{Department of Solid State Physics, Indian Association for the Cultivation of Science, 2A \& 2B Raja S. C. Mullick Road, Jadavpur, Kolkata 700 032, India}
\author{Carlo Meneghini}
\affiliation{Dipartimento di Fisica E. Amaldi, Universita di Roma Tre, Via della vasca navale, 84 I-00146 Roma, Italy}
\author{Indra Dasgupta}
\affiliation{Department of Solid State Physics, Indian Association for the Cultivation of Science, 2A \& 2B Raja S. C. Mullick Road, Jadavpur, Kolkata 700 032, India}
\affiliation{Center for Advanced Materials, Indian Association for the Cultivation of Science, 2A \& 2B Raja S. C. Mullick Road, Jadavpur, Kolkata 700 032, India}
\author{Sudipta Bandyopadhyay}
\affiliation{Department of Physics, University of Calcutta, 92 Acharya Prafulla Chandra Road, Kolkata 700009, West Bengal, India}
\author{Sugata Ray}
\email[Corresponding author:] { mssr@iacs.res.in}
\affiliation{Department of Materials Science, Indian Association for the Cultivation of Science, 2A \& 2B Raja S. C. Mullick Road, Jadavpur, Kolkata 700 032, India}
\affiliation{Center for Advanced Materials, Indian Association for the Cultivation of Science, 2A \& 2B Raja S. C. Mullick Road, Jadavpur, Kolkata 700 032, India}

\date{\today}

\begin{abstract}
Cubic Half-Heusler Cu$_{1-x}$Co$_x$MnSb (0 $\leq$ $x$ $\leq$ 0.1) compounds have been investigated both experimentally and theoretically for their magnetic, transport and electronic properties in search of possible half metallic antiferromagnetism. The systems (Cu,Co)MnSb are of particular interest as the end member alloys CuMnSb and CoMnSb are semi metallic (SM) antiferromagnetic (AFM) and half metallic (HM) ferromagnetic (FM), respectively. Clearly, Co-doping at the Cu-site of CuMnSb introduces changes in the carrier concentration at the Fermi level that may lead to half metallic ground state but there remains a persistent controversy whether the AFM to FM transition occurs simultaneously. Our experimental results reveal that the AFM to FM magnetic transition occurs through a percolation mechanism where Co-substitution gradually suppresses the AFM phase and forces FM polarization around every dopant cobalt. As a result a mixed magnetic phase is realized within this composition range while a nearly HM band structure is developed already at the 10\% Co-doping. Absence of $T$$^2$ dependence in the resistivity variation at low $T$-region serves as an indirect proof of opening up an energy gap at the Fermi surface in one of the spin channels. This is further corroborated by the $ab-initio$ electronic structure calculations that suggests a nearly ferromagnetic half-metallic ground state is stabilized by Sb-$p$ holes produced upon Co doping.
\end{abstract}

\maketitle

PACS number(s): 71.20.Be, 75.50.Ee, 78.70.Dm, 72.25.Ba, 71.15.Mb

\newpage

\section{Introduction}
Heusler and semi Heusler alloys have received considerable interest both theoretically and experimentally as some of them exhibit half metallic (HM) band structure,~\cite{Galanakis1,Galanakis2,Nanda1,Nanda2,Kudrnovsky,Jeong,PRB84} structural similarity to widely used binary semiconductors (e.g. GaAs, InP),~\cite{Galanakis1} and very high Curie temperature ($T$$_C$$ >$300 K).~\cite{PRB84} Generally full Heusler alloys (general formula $X$$_2$$Y$$Z$) are formed in cubic L2$_1$ structure while half Heusler alloys (general formula $X$$Y$$Z$) have cubic C1$_b$ structure.~\cite{Galanakis3} Here $X$ is a higher valent and $Y$ is a lower valent transition metal element and $Z$ is the nonmagnetic $sp$ atom; e.g. $Z$= Ga, Sn, In, and Sb. Most important functional properties of Heusler alloys are connected with their half metallic (HM) band structure.~\cite{Galanakis3,Sasioglu} Based on electronic structure calculations, de Groot $et$ $al$. proposed NiMnSb as possible half metallic ferromagnet (HMFM).~\cite{Groot1} Experimentally, few half heusler compounds e.g. PtMnSb, CoMnSb~\cite{Otto1,Otto2,Galanakis3} and several full Heusler alloys like Co$_2$MnSi,~\cite{Singh} Co$_2$FeSi,~\cite{PRL2013} Co$_2$VAl, Co$_2$VGa~\cite{Kanomata} are found to be HMFM. However, there is also a proposition of half metallic antiferromagnets (HMAFM), also termed as HM fully compensated ferrimagnets, which can offer the unique possibility of net zero magnetic moment within HM band structure.~\cite{Lee,Wurmeh}
\par
Half Heusler alloys of general chemical formulae $X$MnSb become particularly important in this context due to their tunable magnetic states between ferromagnetic and antiferromagnetic.~\cite{Galanakis2,Kudrnovsky,PRB84,Sasioglu,Nature} On the other hand, choice of $X$ and doping at the $X$-site would change the number of valence electron count and therefore the electron DOS at the $E$$_F$, and consequently the magnetic exchange interactions. Interestingly, CuMnSb (Cu:3$d$$^{10}$; nonmagnetic) is the only AFM [Mn-spin structure is shown in Fig. 1(a)]~\cite{Jeong,Endo1} compound within this family which naturally becomes important in this context although it is a semi-metal. On the contrary, CoMnSb (Co:3$d$$^7$; partially filled, weakly magnetic) shows HMFM~\cite{Galanakis3,Duong}[arrangement of Mn spins is shown in Fig. 1(b)]. Therefore Co-doping at the Cu-site of CuMnSb must drive one AFM to FM transition along with one semi-metal to half metal transition. However, it is not clear that whether the electronic mechanism of these transitions is one and the same and therefore they are bound to coincide in $x$ of Cu$_{1-x}$Co$_x$MnSb or not. Existing literature seems to provide highly contradicting views and results in this regard. In 2007, N. P. Duong {\it et al.} experimentally suggested the HMFM nature of all the compounds with $x~\ge$~0.1 ~\cite{Duong} but refrained from commenting about the said transitions. Interestingly, I. Galanakis {\it et al.} theoretically presented a completely different picture and claimed that the system should become ferromagnetic only above 20\% Co-doping while half-metallicity will arise at $x~>$~0.4.~\cite{Galanakis2} More recently, K. Kumar {\it et al.} showed that the system remains largely ferromagnetic above 20\% Co-doping while half-metallicity does not arise even at 80\% Co-doping.~\cite{Kumar} Overall, there had been one consensus that the pure antiferromagnetism of CuMnSb certainly gets lost within 20\% Co-doping while the appearance of half-metallicity is proposed to occur at any doping between 10\% to 80\%.

\par
Therefore, we were curious to probe whether half-metllicity actually occurs at around 10\% and coexist with antiferromagnetism and consequently attempted to explore the range of 0 $<$ $x$ $\leq$ 0.1 of Cu$_{1-x}$Co$_x$MnSb. Consequently, we have synthesized a set of Cu$_{1-x}$Co$_x$MnSb ($x$ = 0.03, 0.05, 0.08, 0.1) compounds, along with the parent CuMnSb one, aiming to keep Co-concentration very low so that AFM state persists in them. Based on detailed magnetic, transport and magneto-transport measurements it is understood that there exists a gradual change over from pure AFM ground state to a coexisting FM/AFM phase within this range of $x$ which is concomitant with the gradual suppression of $T$$^2$-dependency of resistivity [$\rho$($T$)] variation at low temperature upon increasing Co-content. This absence of $T$$^2$-dependence serves as an indirect proof for half-metallicity.~\cite{Majumder,Nigam2,Nigam3}
\section{Methodology}
\subsection{Experimental Techniques}
Polycrystalline samples of CuMnSb and Cu$_{1-x}$Co$_x$MnSb ($x$ = 0.03, 0.05, 0.08, 0.1) have been synthesized by a melt-quenching technique using an arc furnace in Ar-atmosphere. Appropriate stoichiometric amounts of high purity (greater than 99.99\%) constituent metals have been taken for melting. In order to achieve better chemical homogeneity, all the samples have been remelted 4-5 times and finally the molten ingots were annealed at 650$^{\circ}$~C for 5 days in vacuum-sealed quartz tube in order to reduce atomic disorder.~\cite{Otto2} X-ray diffraction (XRD) measurements have been performed at MCX Beam line of ELETTRA (Trieste, Italy) synchrotron radiation facility at room temperature, with $\lambda$=0.827 {\AA}. The XRD data were analyzed by Rietveld method and refinements were carried out by FULLPROF program.~\cite{Carvajal} The exact stoichiometry of all the alloys are additionally verified by inductively coupled plasma-optical emission spectroscopy (ICP-OES) using a Perkin Elmer Optima 2100 DV instrument. To get information on local atomic order/disorder, Mn $K$-edge Extended X-ray Absorption Fine Structure (EXAFS) measurements of two selected samples ($x$=0 and 0.1) have been performed at the XAFS beam line of ELETTRA (Trieste, Italy) synchrotron radiation facility at room temperature in transmission geometry using a double crystal Si (311) monochromator.~\cite{Di Cicco} Data treatment and quantitative analysis of EXAFS were carried out using ARTEMIS.~\cite{Newville} Transport/magnetotransport properties of the alloys have been measured by standard four probe method within a temperature range of 2 K to 300 K in a physical property measurement system (PPMS, Cryogenics). The temperature and magnetic field dependence of magnetization of all these samples were measured in a SQUID magnetometer (Quantum Design, USA).
\subsection{Theoretical Methods}
The electronic structure calculations of Cu$_{1-x}$Co$_x$MnSb have been performed using plane wave basis set and projector augmented wave method (PAW)~\cite{Blochl} in the framework of density functional theory (DFT) as implemented in the Vienna $ab-initio$ simulation package (VASP),~\cite{Kresse} using local density approximation (LDA) to the exchange and correlation part including a Hubbard onsite $d$-$d$ Coulomb interaction at the Mn site. Suitable superstructures were prepared for Co-doped cases and atomic positions were relaxed to minimize the Hellmann-Feynman force with a tolerance factor of 10$^{-2}$ eV/{\AA}. The plane wave energy cut-off is set to 600 eV. For integrations, the Brillouin-zone was sampled with 4$\times$4$\times$4 $k$-mesh. Self consistent calculations have been performed with energy convergence criteria of 10$^{-6}$ eV.
\section{Results and Discussions}
\subsection{Structure from X-ray diffraction and composition verification by ICP-OES}
Fig. [2(a)-(d)] represent the XRD patterns of all the synthesized alloys at room temperature, along with the refined patterns and the difference spectra. The intensity profile observed here confirms single phase with cubic C1$_b$ structure and space group $F$${-43m}$ for all these alloys. Structural refinements of Cu$_{1-x}$Co$_x$MnSb have been carried out assuming that Cu/Co atoms will occupy the (0,0,0) sublattice, while Mn and Sb atoms should occupy the other two sublattices ($\frac{1}{4}$,$\frac{1}{4}$,$\frac{1}{4}$) and ($\frac{3}{4}$,$\frac{3}{4}$,$\frac{3}{4}$) respectively.~\cite{PRB84} The fourth sublattice ($\frac{1}{2}$,$\frac{1}{2}$,$\frac{1}{2}$) is empty, as verified from Rietveld refinements of all the samples. Presence of any Cu/Co and Mn site-disorder between (0,0,0) and ($\frac{1}{4}$,$\frac{1}{4}$,$\frac{1}{4}$) sites has been clearly excluded from the structural refinements of all the samples. For CuMnSb, the refined lattice constant ($a$ = 6.101 {\AA} as obtained after Rietveld refinement of its Synchrotron XRD data, not shown in the present study) is only marginally different from the previously reported value.~\cite{Boeuf} Owing to the size difference between Co and Cu a systematic shift in the diffraction peaks towards higher 2$\theta$ side (see inset to Fig. 2(d)) has been noticed, indicating gradual decrease in lattice constants as well as unit cell volume. Lattice parameters, unit cell volume, $\chi$$^2$ values, and the Mn-Mn distances along with the stoichiometries obtained from the structural refinements of all the samples are listed in Table-I. In order to quantify the exact stoichiometry of all the samples, we did perform ICP-OES analysis which shows that the actual stoichiometries are very close to the desired ones (see Table-I).
\subsection{Local structural measurements}
In addition to structural refinements by XRD and compositional analysis by ICP-OES, we did perform Mn $K$-edge EXAFS measurements on two selected samples [see Fig. 3(a) to (d)] to probe the local structure around Mn. For CuMnSb, Mn $K$-edge EXAFS analysis clearly reveals structurally perfect Mn-site even at the local level, where each Mn sees four Cu as nearest, six Sb as next nearest and twelve Mn as next to next nearest neighbours.~\cite{Jeong}  As Co/Cu have very similar atomic numbers and doping concentration of Co in these alloys is very less, it is quite impossible to distinguish them, in particularly the coordination numbers around Mn. Therefore, to comment on the local atomic order/disorder at the Cu site in Co$_{0.1}$Cu$_{0.9}$MnSb, we did fit Mn $K$-edge EXAFS data by assuming 4 Cu as nearest neighbors around Mn as well as using different Cu/Co-combination as nearest neighbors in order to see whether the fitting improves significantly. The later did not provide better fitting, instead we had arrived with strong correlation among the structural parameters. This comparative study (not shown here) allows us to conclude that the local structure around Mn in 10\%-doped sample is homogeneously preserved without any impurity clustering. Further, the relatively weaker contribution of Mn-Cu2 single scattering in 10\% Co-doped sample compared to the parent one [appearing in the form of weaker oscillation in $k$$^3$ weighted data, shown in Fig. 3 (a) and (c)] also supports the homogeneous substitution of Co in CuMnSb. All the extracted parameters obtained from EXAFS fitting are presented in Table-II.
\subsection{CuMnSb: Magnetization}
An antiferromagnetic peak is present at around 62 K ($T$$_N$), depicted in Fig. 4(a). Irreversibility between zero field cooled (ZFC) and field cooled (FC) magnetization curves starts to appear near around $T$$_N$ and it persists down to 2 K, indicating certain spin disorder in the system. This irreversibility gets largely reduced at the higher magnetic fields but the $T$$_N$ remains same [see inset to Fig. 4(b)], suggesting robust nature of antiferromagnetism. The slight increase in FC data at extremely low temperature might be due to little disorder in spin orientation at the structurally perfect Mn-sublattice ~\cite{Kudrnovsky}, and hence, at reasonably higher field this minor upturn gets largely reduced [inset of Fig. 4(a)]. Further nearly linear $M$-$H$ behavior (very weak hysteretic behavior at $H$ = 0 is again probably due spin-disorder) without any tendency of saturation magnetization at the highest applied field and lowest measuring temperature (shown in Fig. 4(b)] also confirms overall AFM nature of CuMnSb.
\subsection{Magnetization of the doped samples}
ZFC-FC $M$($T$) curves of Cu$_{1-x}$Co$_x$MnSb ($x$ = 0.03, 0.05, 0.08, 0.1) samples at low constant applied fields have been presented in Figs. 5(a) to (d). A gradual shift of the AFM transition temperature ($T$$_N$) towards lower values has been observed with increasing Co-doping. Interestingly, along with the magnetic transition from AFM to paramagnetic (observed for CuMnSb and 3$\%$-doped samples), another FM-like transition starts to develop for higher doping concentrations ($x$=0.05, 0.08, 0.1). This FM-like transition temperature ($T$$_C$) shifts to higher values and also the magnetic moment increases regularly with increasing Co-doping, which indicates regularly increasing magnetic correlation and strong FM exchange. Microscopically a clear correlation can be drawn between $T$$_C$ and interatomic Mn-Mn distance. Smaller the Mn-Mn distance (due to lattice contraction for Co-doping), larger is the strength of FM interactions and as a consequence higher is the $T$$_C$. The constant value of FC-magnetization below $T$$_N$ possibly indicates some FM-like cluster spins appearing in the AFM matrix of CuMnSb upon Co-doping. During ZFC, these FM clusters freeze in random directions, hence do not contribute in a collective way to the net magnetization. While during FC process, these FM clusters tend to align along field and therefore contribute to a higher and constant value of magnetization below $T$$_N$. Interestingly, for 10\% Co-doping the $T$$_N$ (in the form of weak downturn) becomes visible only during ZFC measurement which remains absent in the FC data [see Fig. 5 (d)]. This is due to the fact that the 10\% Co-doped sample is nearly FM along with traces of AFM islands, and therefore the strength of the AFM component in this sample is very weak. So when the sample is cooled in presence of a modest field the large FM part aligns completely along the applied magnetic field producing large magnetization and thereby, masking the contribution of the tiny AFM component. In order to understand the magnetic ground state, $M$($T$) measurements of two higher doped samples ($x$ = 0.08 and 0.1) at different constant higher fields have been performed, depicted in Figs. 5(e) and (f). The AFM peak starts to become weak and finally completely disappears without any ZFC/FC bifurcation at 5000 Oe fields. Also the magnitude of moment increases monotonically with the increase of field and finally reaches close to saturation with decreasing temperature at the maximum applied field. This indicates a closer analogy to those of typical systems, where competing magnetic interactions coexist.  Indeed change in electron concentration (i.e. difference in Cu/Co concentrations) modifies the density of states at the Fermi level and hence affects the exchange interaction between Mn-Mn spins. As a consequence of which, such AFM to FM phase transition could be attributed to the dominance of FM RKKY type of interaction over the AFM superexchange.~\cite{Galanakis2,Sasioglu} In summary, Cu$_{1-x}$Co$_x$MnSb alloys undergo a smooth systematic change over from AFM to competing AFM/FM to a near FM state as a function of $x$, more like through percolative pathways.
Fig. 6 represents the $M$($H$) isotherms at 5 K for the doped samples. Hysteresis behavior has been observed for all the compositions. Quite large coercivity ($H$$_C$ = 2200 Oe) becomes evident for Cu$_{0.97}$Co$_{0.03}$MnSb (see bottom right inset), indicating large anisotropy of the FM-like clusters (developed upon Co-doping), embedded in the host AFM matrix of CuMnSb. But $H$$_C$ reduces to 350 Oe for $x$=0.05. The values of $H$$_C$, as envisioned from the top left inset, become very marginal for the higher doped samples. This reveals a clear change over from disordered FM phase (isolated FM clusters around every dopant Co within AFM host) to a soft FM state (within an overall FM state very little traces of AFM domains) with increasing impurity concentration. Also the magnitude of maximum obtained moment ($M$$_{max}$) increases with doping. It might be assumed here that the FM clusters are growing in volume within the sample with the increase of impurity doping in CuMnSb. This is well consistent with the fact that the smaller volume of FM regions, due to large intrinsic anisotropy, requires higher applied fields for aligning them and as a consequence greater is the coercivity for low doped samples. On the other hand, RKKY-type exchange driven ferromagnetic correlations increase in weightage with increasing dopant concentration, thereby reducing the anisotropy of the isolated FM-like clusters. Thus, for higher doped compositions, larger volume of FM regions could be aligned quite easily along the field even by applying very modest magnetic fields and consequently, coercivity gets largely reduced. The volume fractions of the AFM and FM states have been estimated for the doped samples from their maximum magnetization, ($M$$_{Max}$) values at the highest applied fields and lowest measuring temperatures. From the spin-polarized DFT calculations, it was concluded that the total spin magnetic moment in half metallic ferromagnetic CoMnSb to be 3.00 $\mu$$_B$/f.u.~\cite{Galanakis2,Galanakis3} In the present study, $M$$_{Max}$ for the 3, 5, 8 and 10\% Co-doped samples are 1.16 $\mu$$_B$/f.u, 1.48 $\mu$$_B$/f.u, 2.26 $\mu$$_B$/f.u, and 2.96$\mu$$_B$/f.u respectively. So the volume fractions of the FM phase are about 39, 51, 75, and 98.6\% for $x$ = 0.03, 0.05, 0.08 and 0.1 respectively, while the remaining fractions are the AFM phase.
\subsection{CuMnSb: Resistivity}
The resistivity variation [$\rho$($T$)] of CuMnSb at zero field as well as at an applied field of 5 Tesla in the temperature range 2-300 K has been presented in Fig. 7(a). The residual resistivity ratio (defined as: RRR = $\rho$$_{300K}$/$\rho$$_{2K}$) has been estimated to be 4.89 for CuMnSb. Although there is no experimental evidence of significant atomic disorder due to site-interchange or accidental occupancy to vacant ($\frac{1}{2}$,$\frac{1}{2}$,$\frac{1}{2}$) sublattice, large value of the residual resistivity ($\rho$$_0$) and consequent low RRR value indicate presence of residual spin disorder scattering down to the lowest measuring temperature.~\cite{Otto1,Otto2} In the high temperature limit 220-400 K [inset to Fig. 7(a)], linear $\rho$($T$) variation establishes the dominance of electron phonon scattering. The sharp resistivity drop at the onset of magnetic ordering temperature can be commonly described in terms of freezing of the spin scattering at the ordered phase.~\cite{Jeong} Well below $T$$_N$, resistivity data has been fitted satisfactorily by simple power law dependence: $\rho$($T$) = $\rho$$_0$ + B$T$$^2$ [see Figs. 7(b) and (c)] for both with and without field . The temperature range of observing $T$$^2$ dependence of $\rho$ becomes shorter upon the application of field. The origin of such dependence may possibly be related to the spin-flip scattering of $s$-electrons by spin-density fluctuations of $d$-electrons via an $s$-$d$ exchange interaction.~\cite{Ueda1,Hertel} Also decrease in the coefficient of quadratic $T$-dependence with $H$ (B=0.035 $\mu$$\Omega$-cm-K$^{-2}$ at $H$=0 to B=0.029 $\mu$$\Omega$-cm-K$^{-2}$ at $H$=5 Tesla) supports the fact that external magnetic field helps to quench the spin fluctuations.~\cite{Ueda2}
\subsection{Resistivity variation of the doped samples}
Fig. 8(a) provides a comparative resistivity variation $\rho$($T$) of Cu$_{1-x}$Co$_x$MnSb ($x$ = 0, 0.03, 0.05) samples, indicating good metallic character of these alloys. However, the residual resistivity increases with composition (65 $\mu$$\Omega$-cm for $x$ = 0.03 to 160 $\mu$$\Omega$-cm for $x$ = 0.05). Further slope-change in $\rho$($T$) data near the magnetic transition temperatures indicates the fact that spin ordering actually controls the electronic transport in these compounds. We have fitted the $\rho$($T$) data of $x$=0.03, 0.05 and 0.1 samples using the equation: $\rho$($T$) = $\rho$$_0$ + B$T$$^n$ [see Figs. 8(b), (c) and (d)]. Although in parent CuMnSb, spin-flip scattering of charge carriers causes quadratic $T$ dependence of $\rho$ in low temperature region, ruling out the possibility of half-metallicity,~\cite{Otto1,Otto2} a gradual suppression of this $T$$^2$-dependence (as indicated by $n$-values of the low temperature fitting) occurs with increasing Co-concentration. It was reported earlier that slight modification in spin-down density of states (DOS) at $E$$_F$ can cause significant variations in the transport properties.~\cite{Pierre}  Further recent experimental reports by L. Bainsla $et$. $al$.~\cite{Nigam2} and Lakhan Pal $et$. $al$.~\cite{Nigam3} have claimed that the absence of $T$$^2$-dependency in $\rho$($T$) variation at low temperature serves as an indirect sign of half metallicity. So clear absence of $T$$^2$-dependence might be correlated with gradual reduction in spin-down DOS at $E$$_F$ upon increasing doping percentage. Hence, with Co-doping these systems (3, 5, and 10\%) possibly reach close to desired half-metallicity within mixed magnetic (AFM plus FM) state.
The effect of applied magnetic field on the $\rho$($T$) variation of both 5\%- and 10\%-doped alloys has been presented in Figs. 9(a) and (b). Instead of sharp transition, smooth rounding up has become visible for both the samples near their respective magnetic transition temperatures. Insets to the corresponding figures show the percentage change in resistivity (i.e. Magnetoresistance or $MR$\%= [[$\rho$($H$)-$\rho$(0)]/$\rho$(0)]$\times$100\%) as a function of temperature. In contrast to 5\% Co-doped sample, presence of single minima around $T$$_C$ of Cu$_{0.9}$Co$_{0.1}$MnSb [top left inset of Fig. 9(b)] also supports the absence of AFM phase at higher fields. Actually 10\% Co-doped CuMnSb is nearly FM within which traces of AFM islands remain. So the strength of this AFM phase is very weak and therefore, it gets completely destroyed at the expense of FM alignment under weak magnetic fields [as weak AFM downturn is visible only in ZFC but missing during FC, shown in Fig. 5 (d)]. Therefore, zero field resistivity variation [Fig. 9(b)] fails to probe any slope change in the $\rho$-$T$ variation of Cu$_{0.9}$Co$_{0.1}$MnSb below 8 K ($T$$_N$). Further a distinct change in the nature of resistivity, from metallic to semiconducting-like [from positive to negative temperature dependence of $\rho$, i.e., $d$$\rho$/$d$$T$, the slope of the curve gets changed from positive to negative, shown in bottom right inset to Fig. 9(b)], takes place near around $T$$_C$ of 10\%-doped sample. We therefore comment that this metal-semiconductor transition is intimately connected with magnetic FM-PM phase transition and it can be qualitatively argued that at the onset of magnetic transition, the energy gap of at least one of the spin bands disappears leading to half metallic behavior of resistivity.~\cite{Majumder} Therefore it appears that the emergence of ferromagnetic order is intrinsically connected half-metallic density of states. Although the sign-change of the coefficient of resistivity and its sensitivity to $H$ around $T$$_C$ are quite similar in nature to those observed in CMR manganites,~\cite{AnilKumar} the exchange interactions and underlying microscopic origin of magnetic phase transition are expected to be quite different (charge ordering between Mn$^{3+}$ and Mn$^{4+}$, bringing double exchange-driven FM interactions in manganites while RKKY-type indirect exchange-driven FM correlations between distant Mn atoms appear in our systems). Such anomalous $\rho$($T$) variation could be connected with the spin dependent scattering of charge carriers with localized spins in the FM state of 10\% Co-doped alloy, as suggested from the theoretical description of resistivity and magnetoresistance of the FM metals with localized spins, developed by M Kataoka.~\cite{Kataoka} 
\subsection{Field dependent Magnetoresistance ($MR$) study}
Finally $MR$($H$) variations of Cu$_{1-x}$Co$_x$MnSb ($x$ = 0.05, 0.1) samples at different constant temperatures have been presented in Fig. 9(c) and (d). At 2 K, 5\% Co-doped CuMnSb exhibits positive ${MR}$ (nearly 5\%) in the low field region ($H$ = $\pm$5000 Oe) and then a cross over from positive to negative ${MR}$ has been observed with increasing field (see Fig. 9(c)-i). As discussed in the $M$-$T$ measurements, Cu$_{0.95}$Co$_{0.05}$MnSb has the AFM transition temperature at $\sim$37 K. Hence below $T$$_N$, the signature of positive ${MR}$ appears possibly due to spin fluctuations induced by the applied field on the competing magnetic structure (coexistence of FM and AFM) that causes an increase in resistivity.~\cite{Nigam2,Yoon,Niculescu} This trend remains more or less similar in the $MR$($H$) variation at 20 K, only the contribution of positive $MR$ gets reduced (nearly 0.8-0.9\%, shown in Fig. 9(c)-ii), which is due to the fact that field induced spin fluctuation decreases with increasing temperature as it moves towards AFM to FM transition temperature. Also the $MR$ curves in the AFM region show little field induced irreversibility: the zero field value at the beginning of first field increasing path (path 1) is slightly different from the same as the field is cycled in the decreasing path (path 2) as well as slight mismatch in the $MR$ value at the magnetic field where cross over from positive $MR$ to negative $MR$ is taking place. Field induced irreversibility is often observed in systems where contributions from two competing magnetic phases coexist.~\cite{Nigam4,Nigam5} Hence, the $MR$($H$) data (at both 2 K and 20 K) signify that the 5\% Co-doped alloy contains both FM and AFM phases. Finally in the temperature range $T$$_N$ $<$ $T$ $<$ $T$$_C$, the ${MR}$ is nominally negative, as indicated in Fig. 9(c)-[iii]. Similar kind of $MR$ behavior has also been observed in Cu$_{0.9}$Co$_{0.1}$MnSb, as shown in Figs. 9(d)[i]-[iii]. The only difference here is that the contribution of positive $MR$ at 2 K is very weak (effective only within $H$ = $\pm$1250 Oe with maximum value of around 0.5\%) compared to that of 5\%-doped sample. In the temperature range $T$$_N$ $<$ $T$ $<$ $T$$_C$, Cu$_{0.9}$Co$_{0.1}$MnSb composition shows negative $MR$ (Fig. 9(d)-[ii] and [iii]). Field induced irreversibility is absent in both 12 K and 205 K, while presence of extremely weak such feature at 2 K supports the fact that the strength of AFM phase is quite less in this sample and therefore any cross over from positive to negative $MR$ has not been seen at any temperature except at 2 K. Indeed competing magnetic interactions might be responsible for such kind of $MR$ behavior in all these doped alloys. The most likely scenario is that the FM clusters get embedded in the parent AFM matrix upon Co-doping, and these FM clusters grow in size with increase in doping concentration. External magnetic field increases the effective strength of FM regions in the sample, suppressing spin fluctuation and hence overall resistivity decreases.
\subsection{Electronic structure calculations}
In order to understand our experimental observations, Cu$_{1-x}$Co$_x$MnSb model structures were prepared by taking 2$\times$2$\times$2 supercell of the conventional unit cell of the parent half-Heusler system CuMnSb and introducing one, two and three Co atoms per unit cell to simulate  3\%, 5\% and 10\% Co-doped structures. To check whether all nearest neighbor Mn around Co build FM correlation, three model spin configurations for each of the Cu$_{1-x}$Co$_x$MnSb ($x$=0.03, 0.05 and 0.10) systems were prepared. These structures were (A) pure FM interaction between each Mn spin, (B) AFM configuration as in parent CuMnSb (i.e., each \{111\} plane is ferromagnetic but neighboring \{111\} planes are connected antiferromagnetically), (C) within an overall AFM connectivity (similar to pure CuMnSb) only neighboring Mn atoms around each Co impurity are connected ferromagnetically in order to introduce mixed phase, suggested from experimental scenario (see Fig. 10(a) to (c) for 3, 5 and 10\% Co-doped cases). The corresponding stabilization energies immediately assure us that the model structure (C) has the lowest energy among the three magnetic structures considered in our calculation. The variation of energy for the FM and AFM configurations with respect to the configuration (C) as a function of Co doping is shown in Fig 11. We find for both LDA and LDA+$U$ calculations, the energy difference between the FM and AFM configuration gradually decreases with increasing Co concentration, indicating the stabilization of long-range FM order in the system with increasing Co concentration. The stabilization of the FM correlations upon Co doping may be attributed to the peculiar electronic structure of the parent compound CuMnSb. The total spin polarized DOS for AFM CuMnSb and Cu$_{1-x}$Co$_x$MnSb ($x$=3\%, and 10\%) in their respective lowest energy configuration are shown in Fig 12. The total DOS for CuMnSb with 23 valence electrons is roughly consistent with the nominal ionic formula Cu$^{1+}$Mn$^{2+}$Sb$^{3-}$, albeit with a peculiarity that the system is not insulating but metallic where some Mn-$d$ states in the minority spin channel are occupied at the expense of Sb-$p$ holes in the majority spin channel. The characteristic feature of the total DOS upon Co-doping is the progressive depletion of the minority $d$-states and increase of majority states at the Fermi level with increasing Co concentration [see inset of Fig. 12(a)].
It is interesting to note that the substitution of Co for Cu in Cu$_{1-x}$Co$_x$MnSb reduces the number of valence electrons by (23-2$x$) but the DOS at Fermi level (D($E$$_{F}$)) at the majority spin channel increases with the increasing concentration of Co. This is due to the fact that not only Cu-$d$ but also Co-$d$ states are nearly occupied in both the spin channels at the expense of Sb-$p$ holes admixed with some Co-$d$ states, as can be seen from the plot of Co-$d$, Cu-$d$, and Sb-$p$ partial DOS, shown in Fig. 12(b). The Mn-$d$ states are strongly exchange split and therefore supports the local moments in these systems. The Sb-$p$ holes produced upon Co-doping therefore triggers a competition between RKKY mechanism stabilizing ferromagnetic correlation, and super-exchange promoting antiferromagnetism. As a result, for  small Co concentration ferromagnetic correlations remain weakly coexisting, however upon increasing the concentration of Co even by 10\% superexchange interaction is substantially weakened, stabilizing ferromagnetism. Further strong hybridization between occupied Co-$d$ and unoccupied Mn-$d$ in the minority spin channel progressively depletes the states at the Fermi level in the minority spin channel, paving the way for half-metallic ferromagnetism consistent with the results of low temperature resistivity measurements.
\par
As a consequence of large exchange splitting of Mn $d$-states, Mn atom sustains 4.10-4.18$\mu$$_B$ moment in all these samples ; whereas due to very small exchange splitting of Co 3$d$, Co atom induces a very small negative moment~\cite{Galanakis3} of around 0.8-0.95$\mu$$_B$, while Sb and Cu remain nonmagnetic. The variation of net unit cell moment with Co-doping is shown in Fig. 13. Here theoretical results became comparable with the experimentally obtained maximum moment value at the highest applied field and lowest measuring temperature for all the samples. The total moment gradually rises with Co-content because of increasing FM correlations, and the HM nature comes with nearly 3$\mu$$_B$ moment per formula unit for 10$\%$ Co-doping sample.
\par
\section{Conclusion}
In summary, we have investigated the Cu$_{1-x}$Co$_x$MnSb series in the low doped region (0$<$$x$$\leq$0.1) to verify the possible appearance of half-metallicity in this range where the transformation from AFM to FM gets nearly complete. Upto $x$ $\sim$ 0.03, the system is predominantly in the AFM state, with only a small fraction of FM component being introduced. In the doping region 0.03$<$$x$$<$0.1, both FM and AFM phases coexist; where ferromagnetic $T$$_C$ increases and antiferromagnetic $T$$_N$ gets reduced systematically. But at $x$ = 0.1, the system is nearly in the FM state. Actually Co locally polarizes its surrounding Mn spins ferromagnetically, i.e, Mn-spins around Co align parallel to each other forming FM-like clusters in the AFM matrix. These clusters grow in size via percolation upon increasing the Co-content, as more and more Mn-spins around Co align ferromagnetically which finally drives the system to a nearly FM state. However, even at 10\% doping, traces of AFM island remains. These results are consistent with $ab-initio$ electronic structure calculations which point to the fact that Co-doping introduces Sb-$p$ holes in the system that tend to stabilize ferromagnetism by RKKY mechanism. While for small Co-concentration ferromagnetic correlations are weak, but for 10\% Co-doped CuMnSb ferromagnetic correlations dominate over the antiferromagnetic superexchange, driving the system to a nearly ferromagnetic state. Interestingly, almost simultaneously transition from semi-metallic to half-metallic state takes place indicating 10\% Co-doping as the critical composition in this series.
\par
\section {Acknowledgement:}
AB and AP thank CSIR, India for fellowship. SKN thanks CSIR for funding. SR thanks CSIR for funding (project no. 03(1269)/13/EMR-II) and Indo-Italian POC for support to carry out experiments in Elettra, Italy. Authors also thank TRC-DST of IACS and Centre for Research in Nanoscience and Nanotechnology, University of Calcutta for providing experimental facilities.

\newpage

\begin{table*}[t]
\caption{All the samples are refined within a single phase to achieve reasonable $R$$_{wp}$ and $\chi$$^{2}$. Cubic $F$${-43m}$ space group is taken for all the compositions. A comparative compositional study has been made between X-ray diffraction and ICP-OES analysis on each of these samples.}
\resizebox{18cm}{!}{
\begin{tabular}{| c | c | c | c | c | c | c |}
\hline Sample & $a$({\AA}) & $V$({\AA})$^3$ & $\chi$$^{2}$ & Mn-Mn distance({\AA}) & Stoichiometry from XRD & Stoichiometry from ICP-OES \\\hline
 CuMnSb & 6.101 & 227.118 & 3.75 & 4.314 & Cu$_{1.00}$Mn$_{1.0}$Sb$_{1.0}$ & Cu$_{1.04}$Mn$_{1.11}$Sb$_{1.00}$  \\
 Cu$_{0.97}$Co$_{0.03}$MnSb & 6.098 & 226.76 & 6.03 & 4.312 & Cu$_{0.972}$Co$_{0.028}$Mn$_{1.006}$Sb$_{1.00}$ & Cu$_{0.965}$Co$_{0.035}$Mn$_{1.13}$Sb$_{1.00}$ \\
 Cu$_{0.95}$Co$_{0.05}$MnSb & 6.096 & 226.53 & 3.08 & 4.310 & Cu$_{0.941}$Co$_{0.059}$Mn$_{0.993}$Sb$_{1.00}$ & Cu$_{0.942}$Co$_{0.058}$Mn$_{1.05}$Sb$_{1.00}$ \\
 Cu$_{0.92}$Co$_{0.08}$MnSb & 6.091 & 225.98 & 2.71 & 4.307 & Cu$_{0.921}$Co$_{0.079}$Mn$_{0.991}$Sb$_{1.00}$ & Cu$_{0.916}$Co$_{0.085}$Mn$_{1.072}$Sb$_{1.00}$ \\
 Cu$_{0.9}$Co$_{0.1}$MnSb & 6.088 & 225.64 & 3.01 & 4.305 & Cu$_{0.901}$Co$_{0.099}$Mn$_{0.993}$Sb$_{1.00}$ & Cu$_{0.90}$Co$_{0.10}$Mn$_{1.03}$Sb$_{1.00}$ \\
\hline
\end{tabular}
}
\end{table*}

\begin{table}
\caption{Local structure parameters as obtained from the EXAFS analysis of Mn $K$-edge for both CuMnSb and 10$\%$ Co-doped samples. The coordination numbers around Mn were fixed to their specific values, while even otherwise they did not make the fitting better any more. The fixed or constrained values are labeled by “$\ast$.” The absolute mismatch between experimental data and best fit is $R$$^{2}$=0.014 for both the samples. Further bond distances between Mn and its nearest/next nearest/next to next nearest neighbors have been compared with that obtained from structural refinements.}
\resizebox{8.6cm}{!}{
\begin{tabular}{| c | c | c | c | c | c |}
\hline Sample & Shell & $N$ & $\sigma$$^{2}$ ($\times$ 10$^{2}${\AA}$^2$) & $R$({\AA}) & $R$$_{XRD}$({\AA}) \\\hline
CuMnSb  & Mn-Cu1 & 4.0 $\ast$ & 0.92(6) & 2.637(3) & 2.642 \\
   & Mn-Sb1 & 6.0 $\ast$ & 1.44(9) & 3.002(2) & 3.051 \\
   & Mn-Mn1 & 12.0 $\ast$ & 2.34(7) & 4.303(2) & 4.314 \\
   & Mn-Cu2 & 12.0 $\ast$ & 1.64(6) & 5.025(3) & 5.059 \\\hline
Cu$_{0.9}$Co$_{0.1}$MnSb  & Mn-Cu1 & 4.0 $\ast$ & 0.97(8) & 2.620(4) & 2.636 \\
   & Mn-Sb1 & 6.0 $\ast$ & 1.39(4) & 3.006(2) & 3.044 \\
   & Mn-Mn1 & 12.0 $\ast$ & 2.13(5) & 4.294(2) & 4.305 \\
   & Mn-Cu2 & 12.0 $\ast$ & 1.78(7) & 5.035(4) & 5.048 \\
\hline
\end{tabular}
}
\end{table}

\newpage

\begin{figure}
\resizebox{8.6cm}{!}
{\includegraphics[14pt,174pt][580pt,633pt]{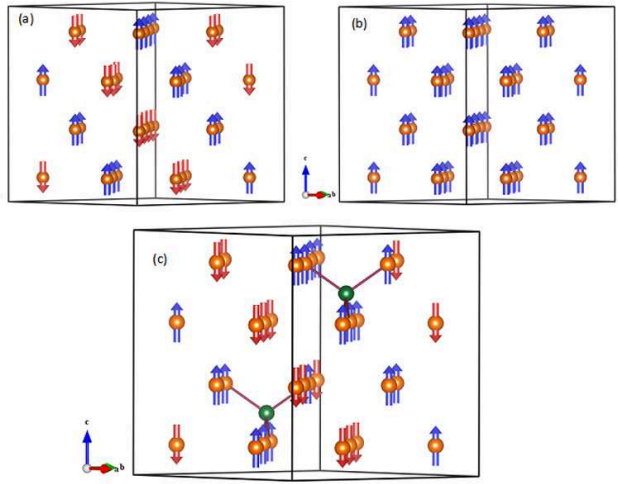}}
\caption{(color online) Arrangement of Mn-spins (golden spheres with red arrows indicating down spin and blue arrows indicating up spin) in pure CuMnSb (a), and pure CoMnSb (b). While for Cu$_{1-x}$Co$_x$MnSb (c), arrangement of Mn-spins to satisfy the hypothesis of mixed magnetic phase. Green spheres indicate the substituted Co atoms.}
\end{figure}

\begin{figure}
\resizebox{8.6cm}{!}
{\includegraphics[8pt,32pt][576pt,624pt]{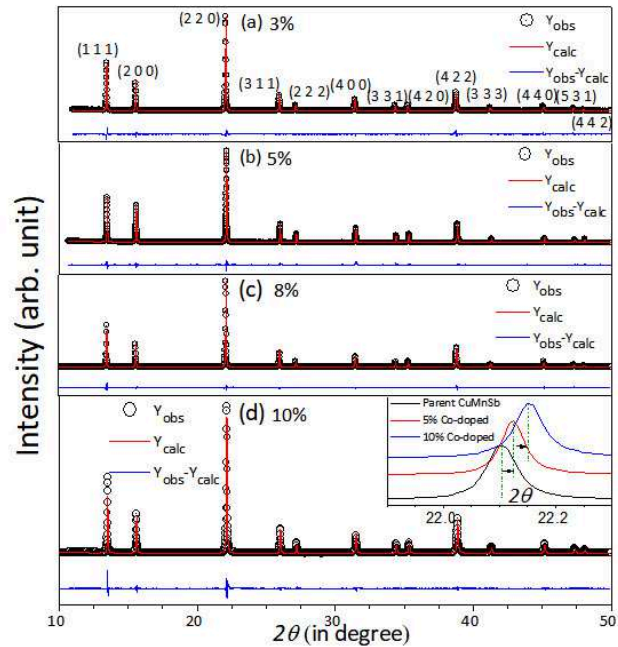}}
\caption{(color online) The refined curves are laid over the experimental data points for all the synthesized samples. Inset to (d) shows the expanded views of the most intense peak of the corresponding samples.}
\end{figure}
\begin{figure}
\centering
\resizebox{8.6cm}{!}
{\includegraphics[6pt,195pt][585pt,642pt]{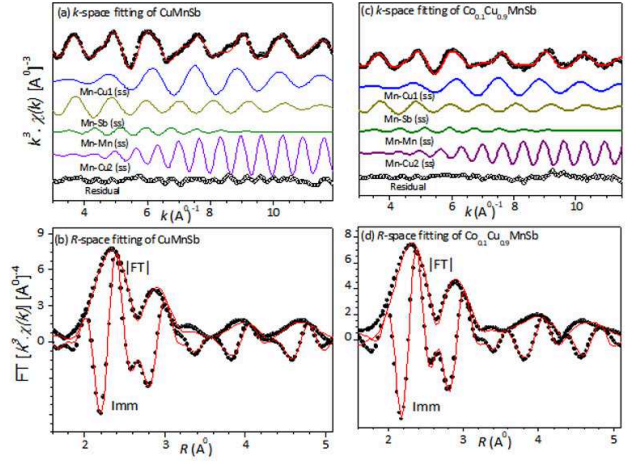}}
\caption{(colour online) (Upper Panel) Mn $K$-edge $k$$^3$ weighted experimental XAFS data (shaded black circles) and respective best fits (red solid line) are presented for both CuMnSb (a) and 10\%-doped sample (c). The partial contributions from individual single scattering paths (solid colored line) and the residual [$k$$^{2}$$\chi$$_{exp}$-$k$$^{2}$$\chi$$_{th}$] (open black dots) are also indicated for both the samples, vertically shifted for clarity. (Lower Panel) Fourier Transform of experimental (shaded black circles) and theoretical (solid red line) curves for CuMnSb (b) and 10\%Co-doped (d) cases; the magnitude ($|FT|$) and imaginary parts (Imm) are plotted in the respective figures.}
\end{figure}
\newpage

\begin{figure*}
{\includegraphics[65pt,347pt][562pt,570pt]{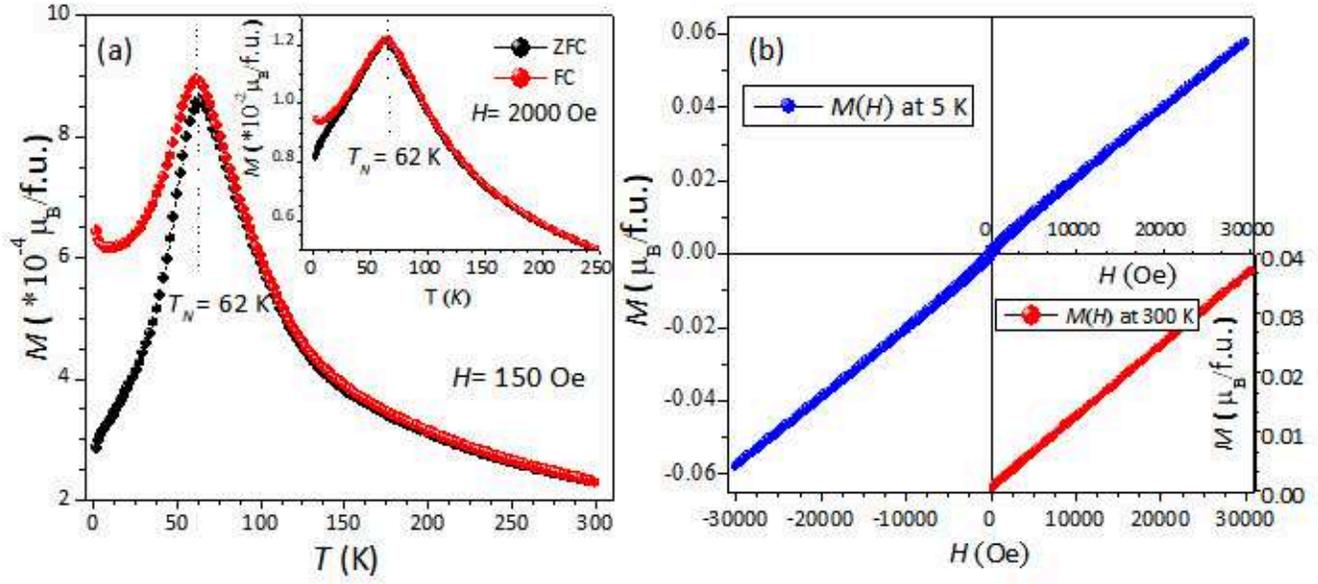}}
\caption{(colour online) (a) $M$($T$) variation for CuMnSb during both Zero Field Cooled (shaded black circles) and Field cooled (shaded red circles) cycles at 150 Oe field; Inset of which additionally shows the $M$($T$) plot at 2000 Oe field. (b) $M$($H$) variation of CuMnSb at 5 K up to ($\pm$)3 Tesla; inset of which shows $M$($H$) curve at 300 K only at the first cycle.}
\end{figure*}

\begin{figure}
\resizebox{8.6cm}{!}
{\includegraphics[15pt,50pt][575pt,773pt]{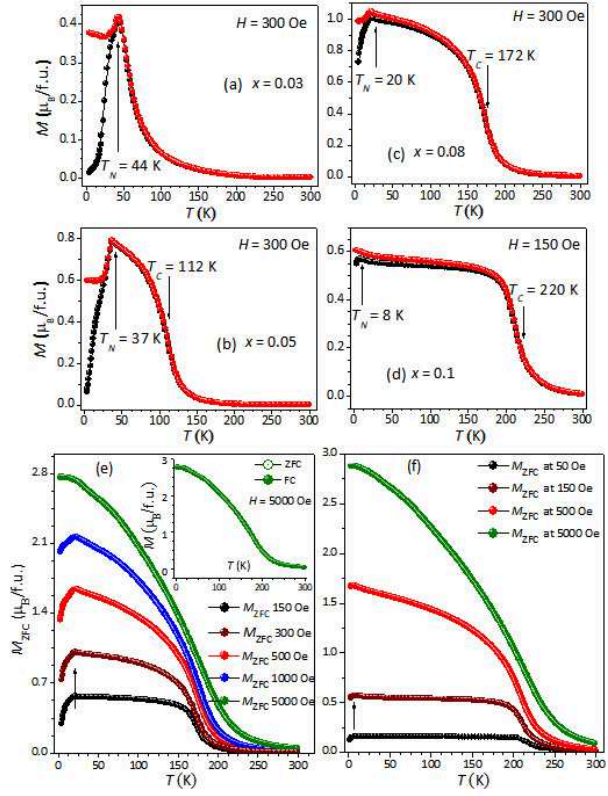}}
\caption{(colour online) (a)-(d) Temperature dependent magnetization [$M$($T$)] variations of all four doped samples at low magnetic field during both ZFC (shaded black circles) and FC (shaded red circles) cycles. Further $M$($T$) variations during ZFC mode of 8\% (e) and 10\%-doped (f) samples have been presented at several constant higher magnetic fields. Inset to (e) shows that ZFC-$M$ and FC-$M$ merge on each other at the highest applied field (5000 Oe) for the 8\%-doped sample.}
\end{figure}

\begin{figure}
\resizebox{8.6cm}{!}
{\includegraphics[42pt,212pt][552pt,608pt]{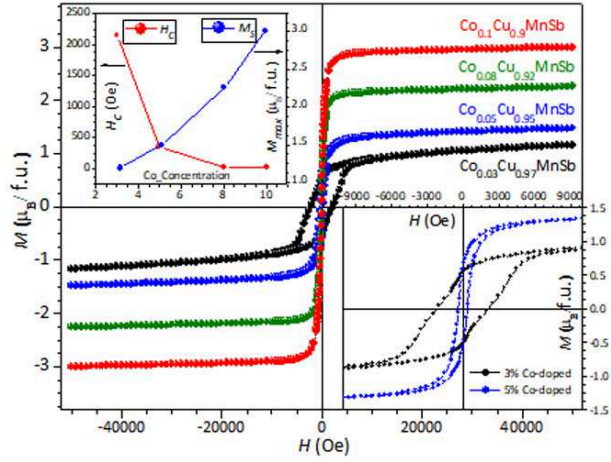}}
\caption{(colour online) (a) The $M$($H$) isotherms at 5 K for the doped samples. Bottom right inset of which shows the enlarged views of the $M$($H$) curves for $x$=0.03 and 0.05 samples in the low field region, while Top left inset shows the variations of maximum magnetization (moment at highest field and lowest measuring temperature) and coercive field ($H$$_C$) as a function of Co-concentration.}
\end{figure}

\begin{figure}
{\includegraphics[73pt,260pt][515pt,524pt]{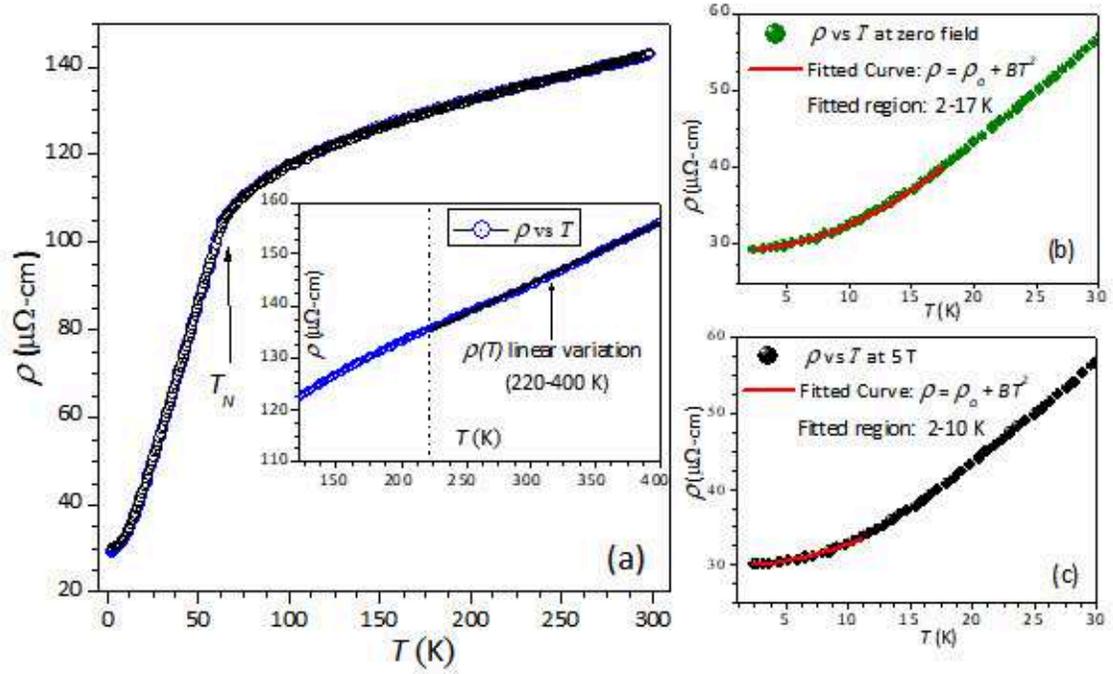}}
\caption{(colour online) (a) Resistivity behavior as a function of temperature [$\rho$($T$)] for CuMnSb. Inset of which shows the linear dependence of $\rho$($T$) (solid black line over experimental data points) in the high temperature region.  Further $T$$^2$-dependency of $\rho$($T$) variation has been shown for zero field (b) and 5 Tesla applied field (c) in the low temperature region.}
\end{figure}

\begin{figure}
{\includegraphics[68pt,311pt][530pt,556pt]{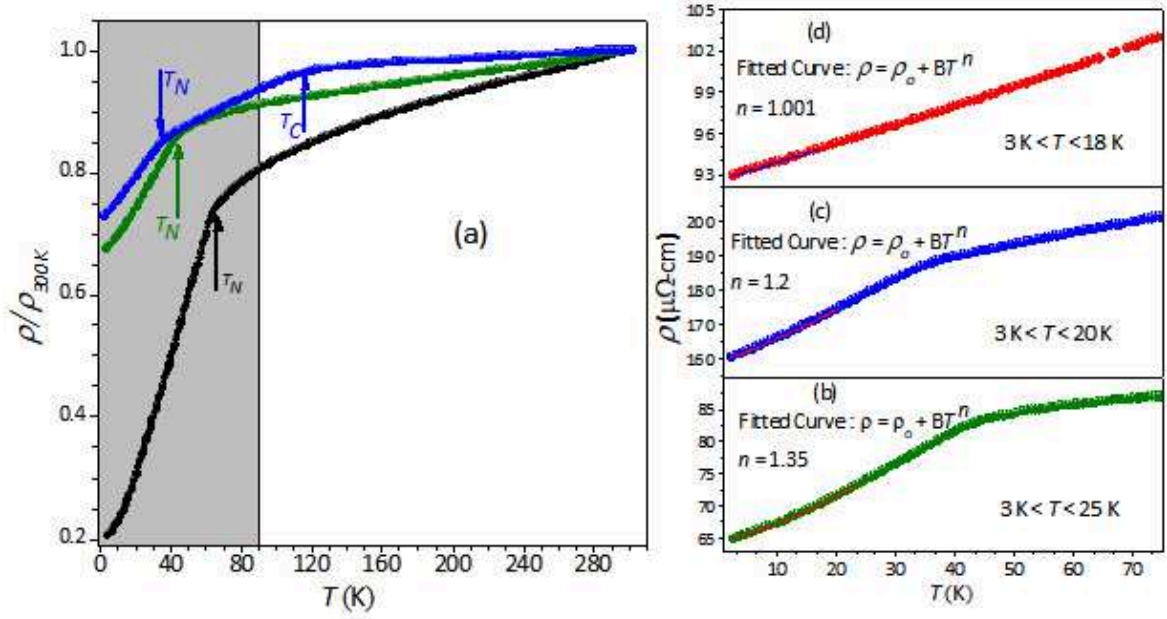}}
\caption{(colour online) (a) Comparative study of normalized zero field resistivity ($\rho$/$\rho$$_{300}$ versus $T$) variation between CuMnSb (shaded balck circles), 3\%Co-doped (shaded green circles) and 5\%Co-doped (shaded blue circles) samples. Further deviation from $T$$^2$-dependence of $\rho$($T$) variation has been shown in a given low temperature range for 3\% (b), 5\% (c) and 10\% (d) doped alloys.}
\end{figure}

\begin{figure}
{\includegraphics[65pt,174pt][552pt,612pt]{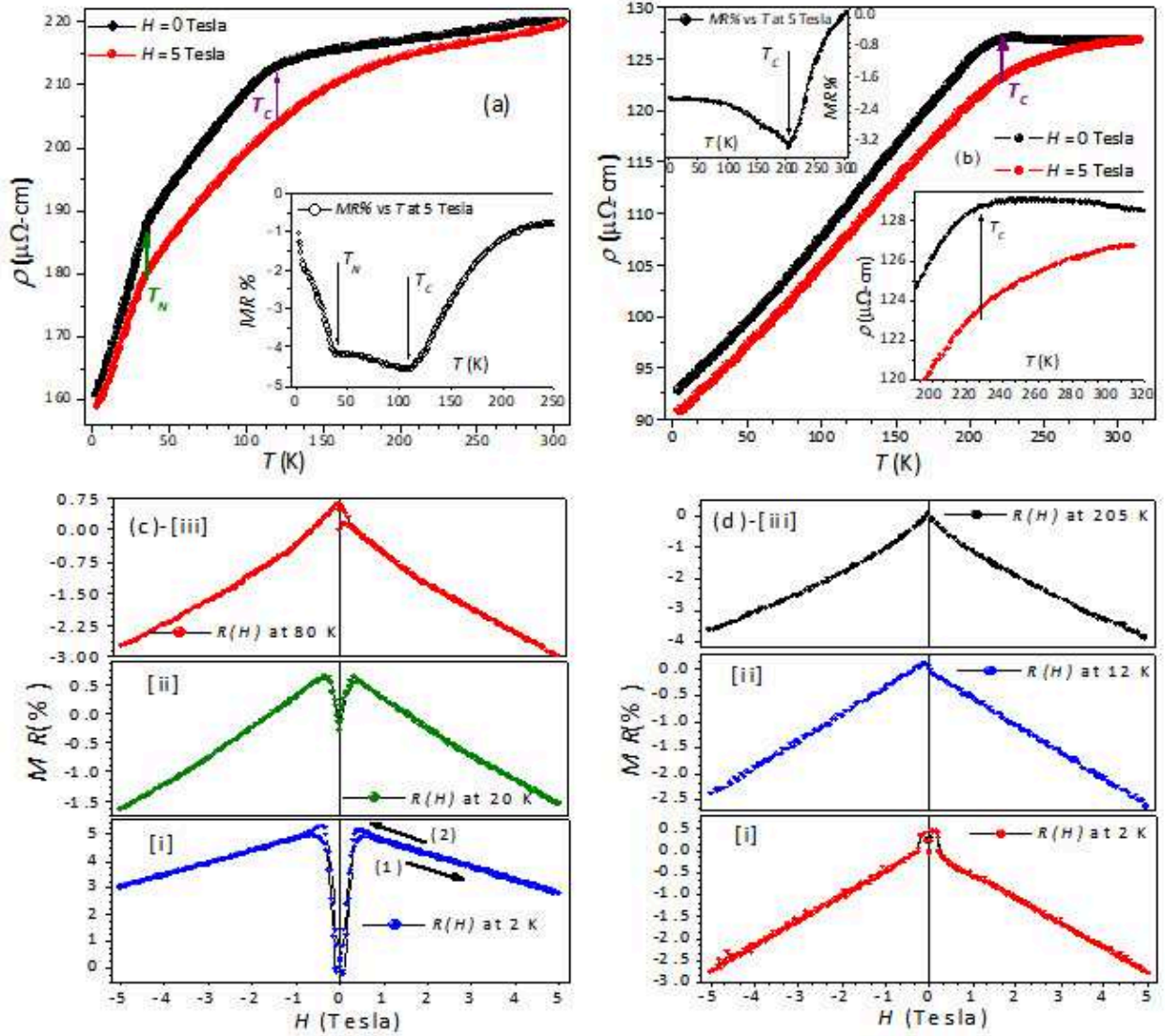}}
\caption{(colour online) $\rho$($T$) variations of 5\%- (a) and 10\%- (b) doped samples at zero field and 5 Tesla applied fields. Inset to (a) shows the $MR$($T$) variation of the corresponding sample, while Top left inset of (b) shows the same for 10\% Co-doped sample and Bottom right inset reveals the enlarged views of the $\rho$($T$) variation at both 0 and 5 Tesla fields in the high temperature region. Further, Magnetoresistance curves ($MR$\% versus $H$) for Cu$_{0.95}$Co$_{0.05}$MnSb ((c)-[i] to [iii]) and Cu$_{0.9}$Co$_{0.1}$MnSb ((d)-[i] to [iii]) alloys have been presented at several constant temperatures.}
\end{figure}

\begin{figure}
\resizebox{15cm}{!}
{\includegraphics[28pt,238pt][577pt,478pt]{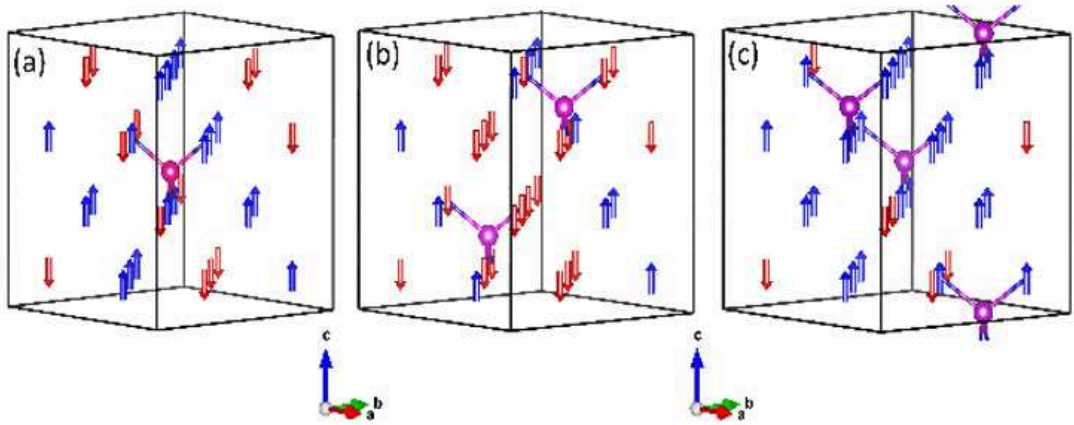}}
\caption{(colour online) Possible spin configurations  retaining ferromagnetic correlation in AFM Cu$_{1-x}$Co$_x$MnSb alloys for 3\% (a), 5\% (b) and 10\% (c) Co-doping. The purple sphere indicates the Co atom while the arrows (blue/red) stand for Mn-spins (up/down).}
\end{figure}

\begin{figure}
\resizebox{12cm}{!}
{\includegraphics[50pt,6pt][575pt,245pt]{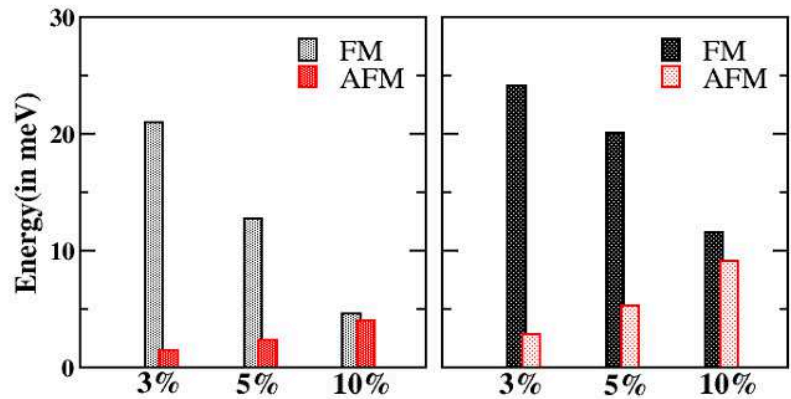}}
\caption{(colour online) Energy difference between FM and AFM configuration with respect to configuration (C)  as a function of Co doping, within  LDA(left panel) and within LDA + $U$ (right panel).}
\end{figure}

\begin{figure}
\resizebox{8.6cm}{!}
{\includegraphics[54pt,45pt][519pt,778pt]{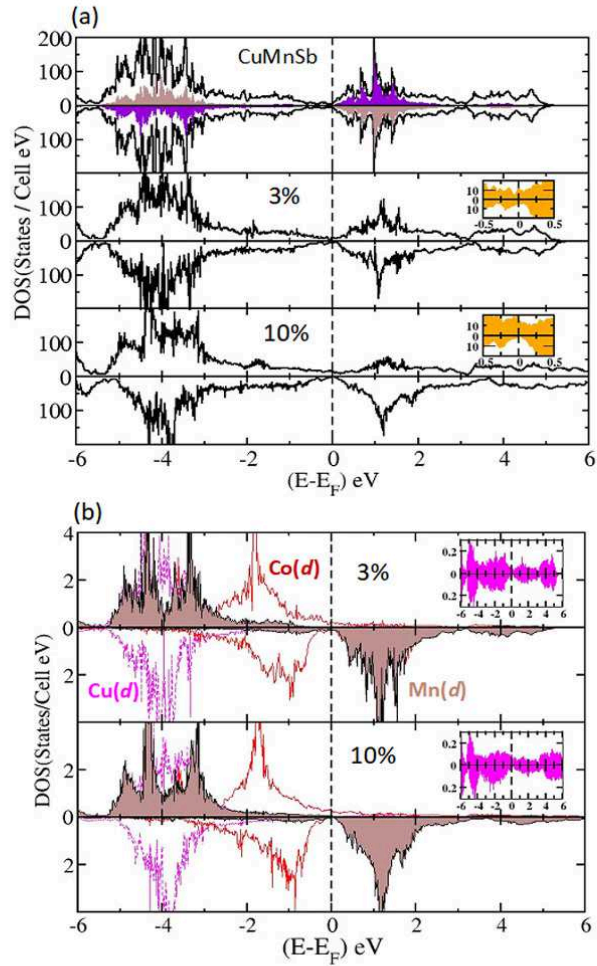}}
\caption{(colour online) (a) Total density of states (TDOS) for the possible ground states of the corresponding systems; Mn 3$d$-projected DOS (filled-in regions) of CuMnSb are also indicated and the insets show the enlarged views of TDOS around the Fermi level for 3\% and 10\%-doped systems. (b) Partial density of states (PDOS) for the possible ground states of 3\%, and 10\% doped structures; insets of which show the PDOS for Sb $p$ states.}
\end{figure}

\begin{figure}
\resizebox{8.6cm}{!}
{\includegraphics[75pt,250pt][512pt,581pt]{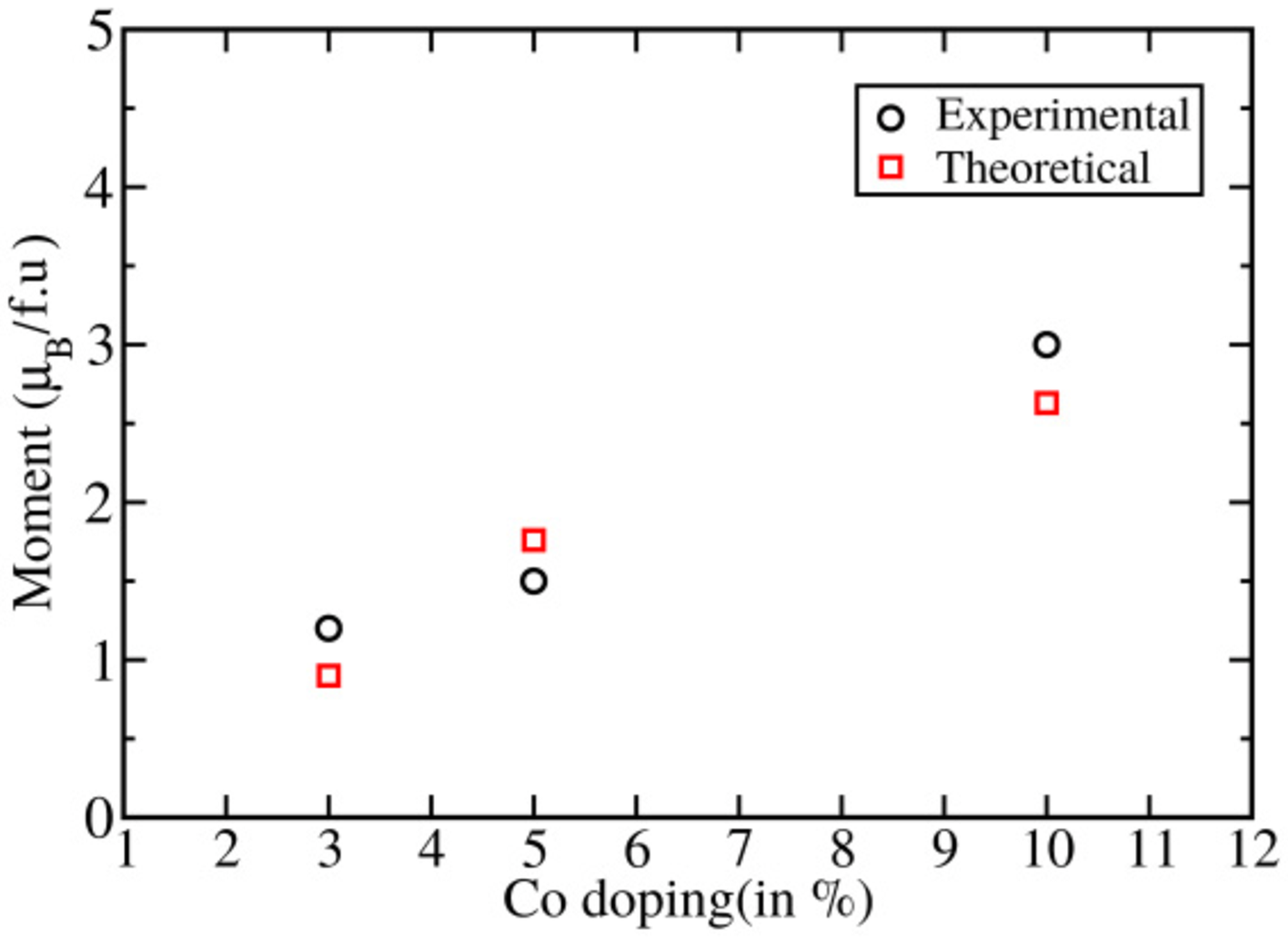}}
\caption{(colour online) Comparison of unit cell moment for 3\%, 5\% and 10\% Co-doped samples.}
\end{figure}


\begin{thebibliography}{99}

\bibitem{Galanakis1} I. Galanakis, K. \"{O}zdo\v{g}an, E. \c{S}a\c{s}io\v{g}lu, and B. Akta\c{s}, Phys. Rev B. {\bf75}, 092407 (2007).

\bibitem{Galanakis2} I. Galanakis, E. \c{S}a\c{s}io\v{g}lu, and K. \"{O}zdo\v{g}an, Phys. Rev B. {\bf77}, 214417 (2008).

\bibitem{Nanda1} B. R. K. Nanda, and I. Dasgupta, J. Phys.: Condens. Matter {\bf15}, 7307 (2003).

\bibitem{Nanda2} B. R. K. Nanda, and I. Dasgupta, J. Phys.: Condens. Matter {\bf17}, 5037 (2005).

\bibitem{Kudrnovsky} J. Kudrnovsk\'{y}, V. Drchal, I. Turek, and P. Weinberger, Phys. Rev. B. {\bf78}, 054441 (2008).

\bibitem{Jeong} T. Jeong, Ruben Weht, and W. E. Pickett, Phys. Rev. B. {\bf71}, 184103 (2005).

\bibitem{PRB84} M. Halder, S. M. Yusuf, A. Kumar, A. K. Nigam, and L. Keller, Phys. Rev B. {\bf84}, 094435 (2011).

\bibitem{Galanakis3} I. Galanakis, P. H. Dederichs, and N. Papanikolaou, Phys. Rev. B. {\bf66}, 134428 (2002).

\bibitem{Sasioglu} E. \c{S}a\c{s}io\v{g}lu, L. M. Sandratskii, and P. Bruno, Phys. Rev. B. {\bf77}, 064417 (2008).

\bibitem{Groot1} R. A. de Groot, F. M. Mueller, P. G. van Engen, and K. H. J. Buschow, Phys. Rev. Lett. {\bf50}, 2024 (1983).

\bibitem{Otto1} M.J. Otto, H. Feil, R.A.M. Van Woerden, J. Wijngaard, P.J. Van Der Valk, C.F. Van Bruggen, and C. Haas, J. Magn. Magn. Mater. {\bf70}, 33 (1987).

\bibitem{Otto2} M. J. Otto, R. A. M. van Woerden, P. J. van der Valk, J. Wijngaard, C. F. van Bruggen, and C. Haas, J. Phys.: Condens. Matter. {\bf1}, 2351 (1989).

\bibitem{Singh} L. J. Singh, Z. H. Barber, Y. Miyoshi, Y. Bugoslavsky, W. R. Branford, and L. F. Cohen, Appl. Phys. Lett. {\bf84}, 2367 (2004).

\bibitem{PRL2013} D. Bombor, C. G. F. Blum, O. Volkonskiy, S. Rodan, S. Wurmehl, C. Hess, and B. B\"{u}chner, Phys. Rev. Lett. {\bf110}, 066601 (2013).

\bibitem{Kanomata} T. Kanomata, Y. Chieda, K. Endo, H. Okada, M. Nagasako, K. Kobayashi, R. Kainuma, R. Y. Umetsu, H. Takahashi, Y. Furutani, H. Nishihara, K. Abe, Y. Miura, and M. Shirai, Phys. Rev. B. {\bf82}, 144415 (2010).

\bibitem{Lee} K. W. Lee, and W. E. Pickett, Phys. Rev. B. {\bf77}, 115101 (2008).

\bibitem{Wurmeh} S. Wurmeh, Hem C Kandpal, G. H. Fecher, and C. Felser, J. Phys.: Condens. Matter. {\bf18}, 6171 (2006).

\bibitem{Nature} A. K. Nayak, M. Nicklas, S. Chadov, P. Khuntia, C. Shekhar, A. Kalache, M. Baenitz, Y. Skourski,
V. K. Guduru, A. Puri, U. Zeitler, J. M. D. Coey, and C. Felser, Nat. Mater. {\bf14}, 679 (2015).

\bibitem{Endo1} K. Endo, J. Phys. Soc. Jpn. {\bf29}, 643 (1970).

\bibitem{Duong} N. P. Duong, L.T. Hung, T.D. Hien, N.P. Thuy, N.T. Trung, and E. Br\"{u}ck, J. Magn. Magn. Mater. {\bf311}, 605 (2007).

\bibitem{Kumar} K. Kumar, A. Dashora, N. L. Heda, H. Sakurai, N. Tsuji, M. Itou, Y. Sakurai, and B. L. Ahuja, J. Phys.: Condens. Matter. {\bf29}, 425805 (2017).

\bibitem{Majumder} S. Majumdar, M. K. Chattopadhyay, V. K. Sharma, K. J. S. Sokhey, S. B. Roy, and P. Chaddah, Phys. Rev. B. {\bf72}, 012417 (2005).

\bibitem{Nigam2} L. Bainsla, M. Manivel Raja, A. K. Nigam, B. S. D. Ch. S. Varaprasad, Y. K. Takahashi, K. G. Suresh, and K. Hono, J. Phys. D: Appl. Phys. {\bf48}, 125002 (2015).

\bibitem{Nigam3} L. Pal, S. Gupta, K. G. Suresh, and A. K. Nigam, J. Appl. Phys. {\bf115}, 17C303 (2014).

\bibitem{Carvajal} J. Rodriguez Carvajal, Physica B {\bf192}, 55 (1993).

\bibitem{Di Cicco} A. Di Cicco, G. Aquilanti, M. Minicucci, E. Principi, N. Novello, A. Cognigni, and L. Olivi, J. Phys. Conf. Ser. {\bf190}, 012043 (2009).

\bibitem{Newville} M. Newville, J. Synchrotron Rad. {\bf8}, 322 (2001).

\bibitem{Blochl} P. E. Blochl, Phys. Rev. B {\bf50}, 17953 (1994).

\bibitem{Kresse} G. Kresse, and J. Hafner, Phys. Rev. B {\bf47}, 558 (1993).

\bibitem{Boeuf} J. Boeuf, C. Pfleiderer, and A. Fai{\ss}t, Phys. Rev B, {\bf74}, 024428 (2006).

\bibitem{Ueda1} K. Ueda, and T. Moriya, J. Phys. Soc. Jpn. {\bf75}, 605 (1975).

\bibitem{Hertel} P. Hertel, J. Appel, and D. Fay, Phys. Rev. B. {\bf22}, 534 (1980).

\bibitem{Ueda2} K. Ueda, Solid State Commun. {\bf19}, 965 (1976).

\bibitem{Pierre} J. Pierre, R. V. Skolozdra, Yu. K. Gorelenko, and M. Kouacou, J. Magn. Magn. Mater. {\bf134}, 95 (1994).

\bibitem{AnilKumar} P. A. Kumar, R. Mathieu, P. Nordblad, S. Ray, Olof Karis, G. Andersson, and D. D. Sarma, Phys. Rev. X. {\bf4}, 011037 (2014).

\bibitem{Kataoka} M. Kataoka, Phys. Rev. B. {\bf63}, 134435 (2001).

\bibitem{Yoon} S. Yoon, and J. G. Booth, Phys. Lett. A. {\bf48}, 381 (1974).

\bibitem{Niculescu} V. Niculescu, T. J. Burch, and J. I. Budnick, J. Magn. Magn. Mater. {\bf39}, 223 (1983).

\bibitem{Nigam4} A. K. Nayak, K. G. Suresh, and A. K. Nigam, Appl. Phys. Lett. {\bf96}, 112503 (2010).

\bibitem{Nigam5} B. Maji, K. G. Suresh, and A. K. Nigam, Europhys. Lett. {\bf91}, 37007 (2010).

\end{thebibliography}
\end{document}